\documentclass[aps,prd,showpacs,twocolumn,floatfix,showkeys]{revtex4}
\usepackage{graphicx}% Include figure files
\usepackage{dcolumn}% Align table columns on decimal point
\usepackage{bm}% bold math
\begin{document}

\title{\Large $\Psi \to N {\bar N} m$}

\author{
T.Barnes$^{a,b}$\footnote{Address from 3 Jan. 2011: U.S. Department of Energy, 
Office of Nuclear Physics.}}

\affiliation{
$^a$Physics Division, Oak Ridge National Laboratory,
Oak Ridge, TN 37831-6373, USA\\
$^b$Department of Physics and Astronomy, University of Tennessee,
Knoxville, TN 37996-1200, USA}

\date{\today}

\begin{abstract}
In this invited talk I discuss two recent applications of charmonium ($\Psi$) decays to $N {\bar N} m$ final states,
where $N$ is a nucleon and $m$ is a light meson.
There are several motivations for studying these decays: 1) They are useful for the study of $N^*$ spectroscopy;
2) they can be used to estimate cross sections for the associated charmonium production processes
$p \bar p \to \Psi m$, which PANDA plans to exploit in searches for charmonium hybrid exotics; and
3) they may allow the direct experimental measurement of $NNm$ (meson-nucleon) strong couplings,
which provide crucial input information for meson exchange models of the $NN$ force. The latter two
topics are considered in this talk, which will also compare results
from a simple hadron pole model of these decays to recent experimental data.
\end{abstract}

\maketitle

\section{Introduction}

$\Psi \to N \bar N m$ charmonium decays have previously attracted interest as an unusual channel
for searches for excited $N^*$ nucleon resonances, for example through sequential decays of the type
$\Psi \to (N^* \bar N + h.c. ) \to N \bar N m$.
Recent theoretical work has suggested two additional areas that may be addressed through studies of
$\Psi \to N \bar N m$ decays, which expands $N^*$ searches to the following set of topics:

\vskip 0.3cm
\noindent
1) Investigations of $N^*$ excitations.
\vskip 0.3cm
\noindent
2) Crossing estimates of cross sections for the PANDA processes $p \bar p \to \Psi m$.
\vskip 0.3cm
\noindent
3) Estimates of meson-nucleon ($NNm$) strong couplings.
\vskip 0.3cm

In this talk I will specialize to the latter two (more recent) topics.
Reviews of the first topic, $N^*$ spectroscopy from charmonium decays,
may be found in the literature \cite{Asner:2008nq,Zou:2008be}.

\section{Estimates of PANDA cross sections from $\Psi \to N \bar N m$ decays}

As the PANDA project \cite{Lutz:2009ff}
plans to produce hybrid charmonium exotics using the associated proton-antiproton
production process $p \bar p \to \Psi m$, there is naturally great interest in the scale of
these cross sections for various choices for the charmonium (or charmonium hybrid) state $\Psi$ and light meson $m$.
Unfortunately, very little is known about these cross sections. The only such reaction that has been
observed experimentally is $p \bar p \to J/\psi \pi^0$, which was studied by E760 and E835 at Fermilab. The data consists
of just two E760 data points \cite{Armstrong:1992ae} for the cross section near the $h_c$ mass
and several uncorrected \cite{Andreotti:2005vu} or unpublished \cite{Joffe:2004ce} E835 points.
Clearly any approach that can give cross section estimates at other energies or
in other channels will be of considerable interest to PANDA.

\begin{figure}
\vskip 0mm
\includegraphics[width=0.7\linewidth]{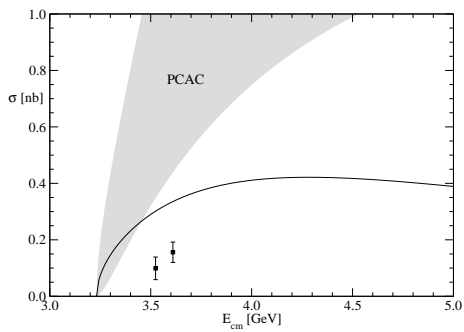}
\vskip -0cm
\caption{The estimated cross section for $p\bar p \to J/\psi \pi^0$
using the constant amplitude formula Eq.3, evaluated by Lundborg {\it et al.} \cite{Lundborg:2005am} (solid line)
A hadron pole model estimate (shaded area labelled PCAC, also from Lundborg {\it et al.})
and the two E760 points 
\cite{Armstrong:1992ae} are also shown.}
\label{f1}
\end{figure}

One interesting possibility is to use BES or CLEO-c data on the three body decays $\Psi \to p \bar p m$ to estimate
the cross sections for $p \bar p \to \Psi m$ at PANDA. This is possible because these two processes are related by crossing,
and as such have a common invariant amplitude.
The $\Psi\to p\bar p m$ differential decay rate (Dalitz plot event density) and
$p \bar p \to \Psi m$ differential cross section are given by

\begin{equation}
d\Gamma_{\Psi\to p\bar p m}
 =\frac{1}{2S_{\Psi}+1}
\frac{1}{(2\pi)^3}\frac{1}{32 M_{\Psi}^3} 
\Big\{ 
\sum |\mathcal{M} |^2
\Big\}
\, dm^2_{m p}\, dm^2_{p\bar p} \ .
\label{eq:width}
\end{equation}
and
\begin{equation} 
\frac{d\sigma}{dt}\bigg|_{p\bar p \to \Psi m} =
\frac{1}{256\pi}\frac{1}{|p_{p\, cm}|^2} s^{-1}
\Big\{
\sum |\mathcal{M}|^2 
\Big\} 
\; .
\label{eq:dsigdt}
\end{equation}
Lundborg {\it et al.} \cite{Lundborg:2005am} used this crossing relation to estimate
near-threshold associated charmonium production cross sections
at PANDA energies from charmonium partial widths under the simplifying assumption of a constant amplitude;
eliminating this constant between the above equations leads to the simple relation
\begin{equation}
\sigma_{p\bar p\to \Psi m}=
4\pi^2 (2S_{\Psi}+1)\,
\frac{M_{\Psi}^3}{A_D}\,
\Gamma_{\Psi\to p\bar p m}\,
\bigg[
\frac{p_{m\, cm}}{p_{p\, cm}} s^{-1}
\bigg].
\label{eq:connect}
\end{equation}
where $A_D$ is the area of the Dalitz plot.
This simple formula gives results for $p\bar p \to J/\psi \pi^0$
that are about a factor of 2-3 larger than the E760 data points,
which is encouraging in view of the simplicity of the approach. This prediction, together with
a range of estimates from a PCAC-like hadron pole model introduced by
Gaillard {\it et al.} \cite{Gaillard:1982zm}
(applied to this process by Lundborg~{\it et al.}~\cite{Lundborg:2005am})
and the two E760 points of Armstrong {\it et al.} \cite{Armstrong:1992ae}
are shown in Fig.~\ref{f1}.

%%%%%%%%%%%%%%%%%%%%%%%%%%%%%%%%%%%%%%%%%%
%%%%%%%%%%%%%%%%%%%%%%%%%%%%%%%%%%%%%%%%%%

Although the constant amplitude results ({\it e.g.} the solid line in Fig.1) are evidently useful
as estimates, one might hope to develop a more realistic model of these
charmonium decays and associated charmonium production reactions. A pole model
for low energy $p\bar p$ associated charmonium production using an effective hadron Lagrangian
was proposed in 1982 by Gaillard {\it et al.} \cite{Gaillard:1982zm}, who considered aspects
of the reaction $p\bar p \to J/\psi \pi^0$ at LEAR.
A similar model has since been applied to $\Psi = \eta_c, J/\psi, \chi_0, \chi_1$ and
$m = \pi^0, \sigma$ and $\omega$ (and flavor partners) in a series of papers by Barnes, Li and
Roberts \cite{Barnes:2006ck,Barnes:2007ub,Barnes:2010yb},
who have calculated $p\bar p \to \Psi \pi^0$ differential and total cross sections \cite{Barnes:2006ck} and
$\Psi \to p\bar p m$ Dalitz plot densities \cite{Barnes:2010yb}. When applied to $\Psi \to p \bar p m$
this model assumes dominance by the Feynman diagrams of Fig.\ref{f3}.

%%%%%%%%%%%%%%%%%%%%%%%%%%%%%%%%%%%%%%%%%%
\begin{figure}[hb]
\vskip  1cm
\includegraphics[width=0.8\linewidth]{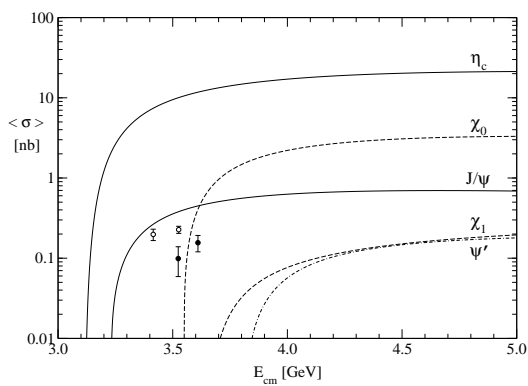}
\vskip -0cm
\caption{Estimates of PANDA $p\bar p \to \Psi \pi^0$ associated charmonium production cross sections
from the hadron pole model \cite{Barnes:2006ck}. The data points shown are all for $p\bar p \to J/\psi \pi^0$,
from E760 (solid) \cite{Armstrong:1992ae} and E835 (open, unpublished) \cite{Joffe:2004ce}.} 
\label{f2}
\end{figure}
%%%%%%%%%%%%%%%%%%%%%%%%%%%%%%%%%%%%%%%%%%

The predictions of this hadron pole model for the PANDA $p\bar p \to \Psi \pi^0$ cross sections
are shown in Fig.3.
Evidently the predicted cross sections for
$\chi_{c0}$ and especially $\eta_c$ production are much larger than the 0.1~[nb] scale of the only experimentally
studied case, $p\bar p \to J/\psi \pi^0$. If correct, this is certainly good news for PANDA. These results suggest
that C=$(+)$ charmonia may have much larger associated production cross sections at low energies
than C=$(-)$ charmonia, perhaps due to the production of C=$(+)$ $c\bar c$ from a $gg$ intermediate state.
Similar estimates of cross sections for associated processes of the type $p\bar p \to \Psi m$ using this model will be
possible given information on the relevant $NNm$ couplings; the use of $\Psi \to N \bar N m$
to estimate these is discussed below.

\section{Estimates of $NNm$ strong couplings from $\Psi \to N \bar N m$}

%%%%%%%%%%%%%%%%%%%%%%%%%%%%%%%%%%%%%%%%%%
\begin{figure}
\vskip -2mm
\includegraphics[width=0.7\linewidth]{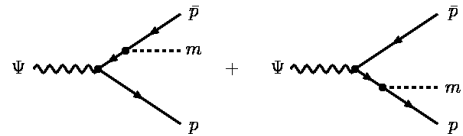}
\vskip -6.0cm
\caption{The Feynman diagrams assumed for $\Psi \to p\bar p m$
in the hadron pole model \cite{Barnes:2010yb}.}
\label{f3}
\end{figure}
%%%%%%%%%%%%%%%%%%%%%%%%%%%%%%%%%%%%%%%%%%

Meson-baryon strong couplings play many important roles in strong interaction physics;
examples include their use in calculating strong decay widths and branching fractions of excited
baryons, and in meson-exchange models of the $NN$ force. Despite their great importance, most of these
couplings (which are actually momentum-dependent strong form factors) are poorly established experimentally.
A rare exception is the pion-nucleon coupling constant, which is $g_{NN\pi} \approx 13$. Although other couplings
such as $NN\omega$ and $NN\rho$ are fitted to data in $NN$ scattering models, it is not clear whether the assumed
meson exchange is actually a realistic description of the short-ranged $NN$ force.

There is an exciting possibility that $\Psi \to NNm$ strong decays may provide a novel set of estimates of these
$NNm$ couplings. To the extent that these decays are well described by the Feynman diagrams of Fig.2,
$NNm$ couplings may be estimated from the ratio $\Gamma(\Psi \to p \bar p m) / \Gamma(\Psi \to p \bar p) $: The
{\it a priori} unknown coupling $g_{NN\Psi}$ cancels out in this branching fraction ratio, leaving an estimate of the
$NNm$ coupling constant (squared). As a test case, an explicit calculation \cite{Barnes:2010yb} of $|g_{NN\pi}|$
using the ratio $\Gamma(J/\psi \to p \bar p \pi^0) / \Gamma(J/\psi \to p \bar p) $ lead to the estimate
\begin{equation}
g_{NN\pi}\bigg|_{J/\psi\to p\bar p \pi^0}  = 13.3 \pm 0.6
\label{eq:g_NNpi_fm_Jpsi_number}
\end{equation}
which is consistent with the standard value of $g_{NN\pi}$. The same calculation using $\psi'$ decays
finds a somewhat smaller estimate of $g_{NN\pi} = 9.9 \pm 0.7$, which suggests some
complications in this approach.

Ideally one should be able to use any initial charmonium state $\Psi$ with allowed
quantum numbers for these coupling constant estimates, so many cross checks should be possible;
the crucial assumption is that the sequential decay Feynman diagrams of Fig.2 are dominant. Unfortunately,
recent CLEO measurements \cite{Onyisi:2010nr}
show that in some $\Psi \to p \bar p \pi^0$ decays the Dalitz plot event distributions
are not consistent with the sequential decay process. In $\chi_{c0} \to p \bar p \pi^0$ in particular,
there is clear evidence for dominance of the Dalitz plot by a low $p\bar p$ invariant mass enhancement,
as has been noted by BES in several other related charmonium decays. In this case the theoretical
$\Psi \to p \bar p m$ decay model must be extended by the inclusion of this low mass $p\bar p$ enhancement;
some possible processes that would lead to this effect have been discussed by
Sibirtsev {\it et al.} \cite{Sibirtsev:2004id}.

One possibility for an $NNm$ coupling measurement in a decay that does not have a strong low mass $p\bar p$ final state
enhancement \cite{:2007dy} is $J/\psi \to p \bar p \omega$, which may allow estimates of the two
$NN\omega$ couplings $g_{\omega}$ and $\kappa_{\omega}$.
Since the $\omega$ plays a very important role in $NN$ force models as the origin of short-ranged $NN$ repulsion,
and the $p \bar p \omega$ Dalitz plot event distribution has been shown to be quite sensitive to the anomalous
coupling $\kappa_{\omega}$ in the hadron pole model \cite{Barnes:2010yb}, it would be very interesting 
to carry out such a study.

\section{Acknowledgments}

I am happy to acknowledge the kind invitation of Hai-Bo Li and the organisers
of Charm 2010 to present this material at Beijing. I also appreciate
the opportunity to discuss charm physics with my colleagues at this meeting, especially Xiaoyan Shen.
This research was supported in part by the U.S. Department of Energy under contract
DE-AC05-00OR22725 at Oak Ridge National Laboratory.
The support of the Department of Physics and Astronomy
of the University of Tennessee is also gratefully acknowledged.

\end{document}